\documentclass{optica-article}

\journal{opticajournal} 

\articletype{Research Article}

\usepackage{hyperref}
\hypersetup{pdfstartview={FitH},pdfpagemode={UseNone},
            colorlinks,linkcolor=blue, citecolor=blue, urlcolor=blue,
            bookmarksopen=true, pdfnewwindow=true}
\usepackage[all]{hypcap}

\usepackage{lineno}

\usepackage[
    style=phys,
    natbib=true,
    doi=true,
    url=true,
    biblabel=brackets,
    sorting=none,
    maxbibnames=99,
    giveninits=true,
    labeldate=year
]{biblatex}

\DeclareSourcemap{
  \maps[datatype=bibtex]{
  \map{
      \step[ 
        fieldsource=url,
        match=\regexp{https://doi.org/(.+)},
        fieldtarget=doi,
      ]
    \step[ 
      fieldsource=doi,
      match=\regexp{https://doi.org/(.+)},
      replace=\regexp{$1}
    ]
    }
    \map{
      \step[fieldsource=doi,final]
      \step[fieldset=url,null]
      \step[fieldset=urldate, null]
    }  
    \map{
      \step[fieldset=month,null]
    }  
  }
}

\DeclareRedundantLanguages{en}{english,american,british,
  canadian,australian,newzealand,USenglish,UKenglish}
\DeclareRedundantLanguages{English}{english,american,british,
  canadian,australian,newzealand,USenglish,UKenglish}

\usepackage{physics}

\begin{document}
\sloppy

\title{Quantum Phase Gradient Imaging Using a Nonlocal Metasurface System}

\author{Jinliang Ren\authormark{1}, Jinyong Ma\authormark{1,2}, Katsuya Tanaka\authormark{3}, Lukas~Wesemann\authormark{4}, Ann Roberts\authormark{4}, Frank Setzpfandt\authormark{5,6}, Andrey~A.~Sukhorukov\authormark{1,*}}

\address{\authormark{1}ARC Centre of Excellence for Transformative Meta-Optical Systems (TMOS), Department of Electronic Materials Engineering, Research School of Physics, The~Australian National University, Canberra, ACT 2601, Australia\\
\authormark{2}{Institute of Quantum Precision Measurement, State Key Laboratory of Radio Frequency Heterogeneous Integration, College of Physics and Optoelectronic Engineering, Shenzhen University, Shenzhen 518060, P. R. China}\\
\authormark{3}Institute of Solid State Physics, Abbe Center of Photonics, Friedrich-Schiller-University Jena, Helmholtzweg 5, 07743 Jena, Germany\\
\authormark{4}TMOS, Department of Electrical and Electronic Engineering, School of Physics, University of Melbourne, Melbourne, Victoria 3010, Australia\\
\authormark{5}Institute of Applied Physics, Abbe Center of Photonics, Friedrich Schiller University Jena, Jena 07745, Germany\\
\authormark{6}Fraunhofer Institute for Applied Optics and Precision Engineering IOF, Albert-Einstein-Straße 7, 07745 Jena, Germany
}

\authormark{*}Corresponding Author: \email{Andrey.Sukhorukov@anu.edu.au}



\begin{abstract*} 

Quantum phase imaging enables the analysis of transparent samples with thickness and refractive index variations in scenarios requiring precise measurements under low-light conditions. 
Here, we present a compact quantum phase-gradient imaging system integrating a lithium niobate (LiNbO$_3$) metasurface for generating spatially entangled photon pairs and a silicon (Si) metasurface for phase gradient extraction. By leveraging nonlocal resonances, the LiNbO$_3$ metasurface enables efficient spontaneous parametric down-conversion (SPDC) with all-optically angularly 
tunable emission, while the Si metasurface provides a nearly linear optical transfer function (OTF) that differentiates the photon wavefunction and extracts phase gradients.
Experimental proof-of-concept results demonstrate the imaging of up to 25~rad/mm phase gradients, achieving 
89$\%$ similarity with the reference values.
The pixel resolution of the system can be potentially enhanced by orders of magnitude by increasing the metasurface dimensions and resonance quality factor.
Our work showcases the application of metasurfaces in both generating and detecting quantum states and establishes a new paradigm for portable quantum phase-gradient imaging, with potential applications in quantum sensing, microscopy, and LiDAR technology.


\end{abstract*}

\section{Introduction}

Quantum imaging utilizing entangled photon pairs presents significant theoretical and practical advantages over classical imaging systems~\cite{Pittman:1995-3429:PRA, shihQuantum2007c, Padgett:2017-20160233:PTRSA, moreauImaging2019, gilabertebassetPerspectives2019,Cameron:2024-33001:JPPH,Defienne:2024-1024:NPHOT,Forbes:2025-174:CPH}. By exploiting quantum entanglement, image reconstruction can be performed through photon coincidence measurements, providing enhanced signal-to-noise ratio (SNR)~\cite{bridaExperimental2010a,samantarayRealization2017,PhysRevD.23.1693}, operation at ultra-low photon fluxes~\cite{gilabertebassetPerspectives2019}, and improved security against eavesdropping~\cite{rusca2024QKDarxiv}. These advantages arise from the intrinsic correlations of entangled photons, which allow the extraction of information with greater precision than classical approaches for the same photon illumination.

Building on this foundation, quantum phase imaging~\cite{Lemos:2014-409:NAT, Paterova:2018-25008:QST, Giovannetti_2011, Ortolano2023} combines the strengths of quantum imaging with phase-sensitive detection~\cite{Nguyen2022,Fienup2013,Park2018,ZERNIKE1942974}, enabling the analysis of transparent samples through thickness and refractive index variations. This capability is particularly valuable in scenarios requiring non-destructive and highly precise measurements under low-light conditions. Conventionally, the measurement of spatial phase profiles and phase gradients relied on bulky optical setups. Whereas recent advances in optical metasurfaces and thin-film structures have provided a compact alternative by realizing first-order differentiation of the optical wavefield and other mathematical operations, enabling phase-gradient extraction and edge enhancement under classical light illumination~\cite{Silva:Science:2014, Zhu2017,  Kwon:NaturePhotonics:2020, Wesemann:2021-98:LSA, Wesemann:APR:2021, Ji2022}, the latter approaches remained unexplored in the realm of quantum phase imaging.

In parallel, nonlinear flat optics has emerged as a versatile platform for generating entangled photons through spontaneous parametric down-conversion (SPDC), see the reviews~\cite{Wang:2022-38:PT, kanAdvances2023, maEngineering2024} and references therein.
Resonance-enhanced nonlinear metasurfaces can significantly boost photon-pair generation~\cite{Santiago-Cruz:2021-4423:NANL} and enable precise engineering of quantum states across frequency~\cite{Santiago-Cruz:2022-991:SCI,Noh:2025-371:LSA}, momentum~\cite{Zhang:2022-eabq4240:SCA}, and polarization~\cite{Ma:2023-8091:NANL, Jia:2025-eads3576:SCA, Noh:2024-15356:NANL, Ma:2025-eadu4133:SCA} degrees of freedom.
Such metasurfaces also offer greater environmental stability compared to conventional bulk crystals, making them attractive for free-space quantum applications. Recently, quantum ghost imaging of amplitude objects was achieved with a metasurface quantum source~\cite{Ma:2025-2:ELI}, however that scheme was insensitive to the phase variations in transparent objects.




In this work, we suggest and demonstrate experimentally a novel ultra-compact approach to phase-gradient imaging through an integrated platform that uses metasurfaces for both the tasks of quantum light generation and phase-gradient extraction. 
Spatially entangled photon pairs are generated by a nonlinear lithium niobate (LiNbO$_3$) metasurface exhibiting nonlocal resonances, where the emission angle is narrow in one direction but broad in the orthogonal direction~\cite{Weissflog:2024-3563:NANP}. 
For phase-gradient extraction, a silicon (Si) metasurface with angle-sensitive nonlocal resonances is designed
that allows first-order differentiation of the single-photon wavefunction and direct retrieval of the phase gradient through two-photon correlation measurements. 
The result is a compact and all-optically tunable system that opens promising opportunities for high-sensitivity, low-light applications in areas such as biomedical imaging, remote sensing, and secure communications.

\section{Results}
\subsection{Concept of phase gradient imaging with single photons} \label{sec:concept}

\begin{figure}[t!]
    \centering
    \includegraphics[width=0.7\linewidth]{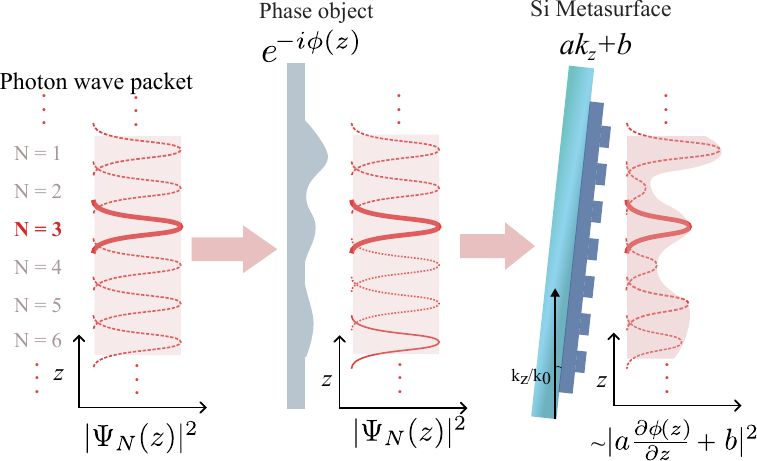}
    \caption{\textbf{Schematic diagram of phase-gradient imaging using quantum light}.
    (Left)~Single-photon wave packets $\Psi_N(z)$ are prepared at different spatial positions labeled $N=1,2,3,\ldots$  in the transverse direction $z$. Each wavepacket has a finite width with a Gaussian-like shape of the wavefunction norm $|\Psi_N(z)|^2$, as indicated by lines.
    (Middle)~
    After propagating through a phase object, the wavepackets acquire a spatially varying phase $-\phi(z)$, while their probability amplitude remains unchanged. 
    (Right)~The Si metasurface imposes a linear optical transfer function ($a k_z + b$) on the wavefunction in $k_z$-space, equivalent to taking a spatial derivative of phase in real space, see an expression under the plot. Consequently, the phase variation encoded in the quantum state is converted into measurable photon rate modulation, allowing direct extraction of the phase gradient at the single-photon level.}
    \label{fig:Fig1_0}
\end{figure}

\begin{figure}[!ht]
    \centering
    \includegraphics[width=1\linewidth]{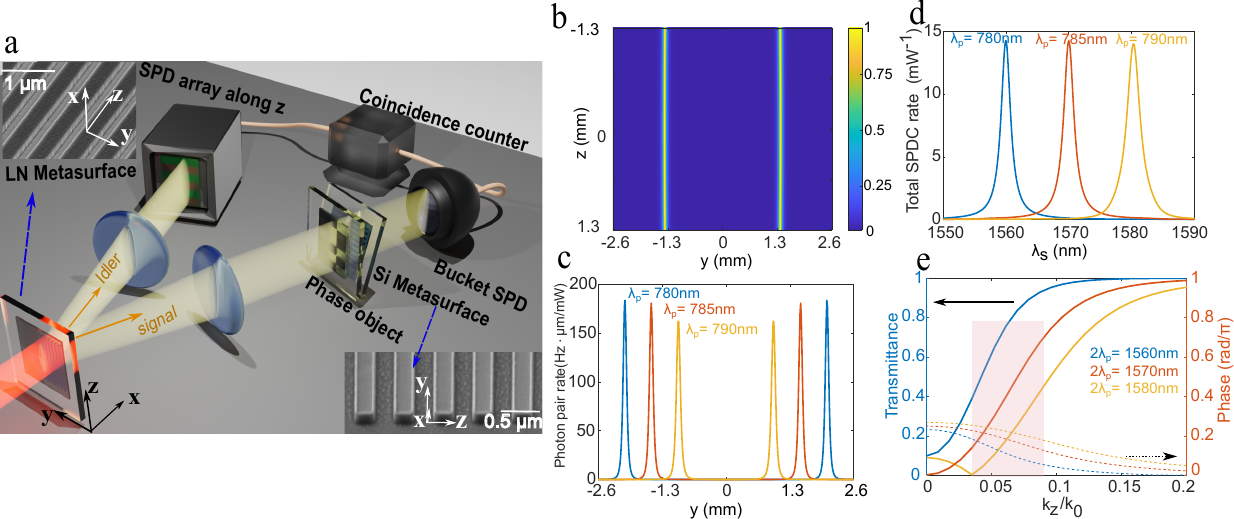}
     \caption{Concept of metasurface system for quantum phase-gradient imaging with simulated features. \textbf{a}. The sketch of the setup. The transverse directions are $z$ and $y$.  The SEM image of the LiNbO$_3$ metasurface is shown in the top-left corner, and that of the Si metasurface is shown at the bottom-right corner. The phase object is prepared with a spatial light modulator (SLM). The Si metasurface is placed immediately after the object with a slight tilt to produce a linear OTF. ~\textbf{b}. The photon-pair emission pattern from LiNbO$_3$ with anti-symmetric uniform $z$-direction distribution and narrow $y$-direction distribution. The wavefield propagation direction is set to the $x$-direction.  ~\textbf{c}. The tunable emission along the $y$-direction. Tuning the pump wavelength shifts the SPDC emission along the $y$-axis, which is used for object imaging in this direction.  ~\textbf{d}. The spectrum of the metasurface-based SPDC process for three selected pump wavelengths. The bandwidth of the generated photons is around 3~nm. ~\textbf{e}. The optical transfer function of the Si metasurface: amplitude (solid lines) and phase (dashed lines) vs. the wavevector component along $z$ normalised to the free-space photon wavenumber.
     The shaded box marks the wavevector range where transmission amplitude dependence is linear at all the three photon wavelengths.
     }
    \label{fig:Concept}
\end{figure}

We develop a scheme for phase-gradient imaging by harnessing a general framework of quantum ghost imaging, which offers a range of practical advantages, including reduced noise at low light levels~\cite{Pittman:1995-3429:PRA, Padgett:2017-20160233:PTRSA, Forbes:2025-174:CPH}. 
In this approach, single-photon wave packets that are transmitted through an object are detected by a simple bucket detector that only counts the photons but does not distinguish their position or momentum state. Then, an object can be imaged by preparing the incident photons with wavefunctions $\Psi_N(z)$ centered at different spatial positions $z_N$, as sketched on the left side of Fig.~\ref{fig:Fig1_0}. 
%
A characteristic single-photon wavefunction can be approximated with a Gaussian shape of the width $\sigma_z$,
\begin{equation}
    \Psi_N(z) =\frac{1}{(2\pi\sigma_z^2)^{1/4}} \exp\left({-\frac{(z-z_N)^2}{4\sigma_z^2}}\right).
\end{equation}
After the photons propagate through the phase object, their wavefunctions exhibit the corresponding phase modulations $O(z)=e^{-i\phi(z)}$, however the norm of the wavefunction $|\Psi_N(z)|^2$ and thereby the photon rate do not change. 

To detect the influence of the phase object, we suggest positioning in the photon path a specially designed Si metasurface mounted at a slight tilt, which features a linear optical transfer function (OTF) 
\begin{equation}\label{eq:LOTF}
    H(k_{z}) = a k_{z}+b
\end{equation}
%
in the transverse momentum space $k_z$, where $a$ and $b$ are constant coefficients. As established in the classical optics regime, such a metasurface can effectively perform the wavefront differentiation in the $z$ direction~\cite{Wesemann:APR:2021,Bykov:2018：optica,Wang:2020:ACSPh}. However, the operation needs to be specially analyzed and optimized in the quantum regime.
The detectable photon rate can be estimated as (see Supplementary Sec.~S2 for the derivation details)
\begin{equation}\label{eq:Demo}
\begin{split}
        C(z_N)=&\int dz \bigg|\mathcal{F}_{z}^{-1}\bigg[H(k_{z})\mathcal{F}_{z}[O(z)\Psi_N(z)]\bigg]\bigg|^2 
        \propto \left[\frac{a^2}{4\sigma_z^2}+\left(a\frac{\partial\phi(z)}{\partial z}+b\right)^2\right]_{z=z_N} ,
\end{split}
\end{equation}
where $\mathcal{F}_{z}$ indicates the Fourier transform along the $z$-direction. 
The 
nonzero constant $b$ is related to the metasurface tilt, which introduces an asymmetry into the phase gradient term and allows the distinction between positive and negative phase gradients for the unambiguous reconstruction of phase modulations. 

We note that to perform accurate phase gradient extraction, the Fourier spectrum of the photon wavefunction must lie entirely within an approximately linear region of the OTF of the width $\Delta k_z$, i.e. we require that 
$\sigma_z > 1 /\Delta k_{z}$. 
This condition sets the lower limit on the spatial resolution, since the object phase gradient is effectively averaged over the $\sigma_z$ width of the photon wavepacket. This is consistent with the resolution limit in the classical case, being also inversely proportional to $\Delta k$ according to the Fourier transform properties.




\subsection{Nonlinear metasurface for photon-pair generation}

In order to prepare the required single-photon states $\Psi_N(z)$ for object imaging, we employ the momentum correlations between two photons generated through the process of spontaneous parametric down-conversion (SPDC) according to the quantum ghost imaging scheme.
In traditional setups, bulk nonlinear crystals were used for generation of entangled photons through SPDC~\cite{Padgett:2017-20160233:PTRSA, Forbes:2025-174:CPH}. Instead, we employ a photon-pair source based on a compact nonlinear metasurface that was recently used to realize ghost imaging of amplitude objects, offering a potential to increase the number of resolution cells by several orders of magnitude~\cite{Ma:2025-2:ELI}.

Accordingly, we propose a complete metasurface-based quantum phase gradient imaging system, as illustrated in Fig.~\ref{fig:Concept}a.
%
Pairs of photons 
are generated through SPDC from a metasurface incorporating a quadratically nonlinear LiNbO$_3$ layer. 
The LiNbO$_3$ metasurface is designed by leveraging nonlocal guided-mode resonances (GMRs) in a lithium niobate thin film, where a SiO$_2$ grating deposited on top mediates resonant coupling between the free-space radiation and the guided modes~\cite{Zhang:2022-eabq4240:SCA}. 
A scanning electron microscope (SEM) image of the fabricated LiNbO$_3$ metasurface is shown in Fig.~\ref{fig:Concept}a. It consists of a sub-wavelength-scale silica grating with a thickness of 200~nm on top of a 300~nm thick $x$-cut LiNbO$_3$ thin film. The grating period was chosen to be 900~nm with a width of 500~nm. The simulation details for the design can be found in Supplementary Sec.~S1.1.
The design ensures that the nonlinear interactions are resonantly enhanced at the photon wavelengths, increasing the signal and idler generation through SPDC.

Due to the conservation of transverse momentum, the signal and idler photons are emitted at opposite angles, and we separately focus them by lenses. 
In the image plane of idler photons, we position a 1D array of single-photon detectors along the $z$ direction, whereas a phase object is placed in the image plane of signal photons. In these planes, the photon emission is broad in $z$, but narrow in $y$ direction, as shown in Fig.~\ref{fig:Concept}b. The photons are anti-correlated in $z$, allowing for ghost imaging, since a detection of idler photon at a particular $z_i$ position heralds the signal photon incidence on the object at the position $z_s \simeq - z_i$~\cite{Ma:2025-2:ELI}. We can thereby perform phase gradient imaging according to the ghost imaging concept discussed in the previous Sec.~\ref{sec:concept} and sketched in Fig.~\ref{fig:Fig1_0}, with $z_N \simeq - z_i$.

In the orthogonal $y$-direction, we perform the imaging by all-optically scanning the photon positions by tuning the pump wavelength~\cite{Ma:2025-2:ELI,Weissflog:2024-3563:NANP}. 
The resonant photon wavelength shifts almost linearly with the incident angle, and this angular dispersion enables a linearly tunable emission angle for the generated photon pairs, ranging from $\lambda_{s,i} = 1560$ to 1580~nm, 
for the pump laser wavelengths from 780 to 790~nm, as illustrated in Fig.~\ref{fig:Concept}c. Importantly, the photon rate is consistently enhanced across all these wavelengths, with the photon bandwidth indicating the metasurface quality factor $Q_{LN}\simeq 500$ according to Fig.~\ref{fig:Concept}d.
The resolution along the $y$-axis is further influenced by the spectral bandwidth (Fig.~\ref{fig:Concept}d) used in the scanning protocol.
The minimum resolvable pixel size along $y$-direction is determined by the condition that cross-talk is small, ensuring reliable phase-gradient extraction. In Sec.~\ref{sec:resolution}, we evaluate these resolution limits through simulations under experimental conditions. 

Overall, the spatial correlation between the paths of signal and idler photons governs the coincidence at each pixel on the 1D detector array along the $z$-direction, with the $y$-coordinate determined by the pump wavelength.
In this way, we can acquire a two-dimensional spatial profile of the object phase-gradient $\partial \phi/\partial z$.


\subsection{Design of Si metasurface for phase gradient imaging}

We now discuss the design principles of a silicon metasurface that is placed after the phase object to facilitate the gradient imaging, as sketched in Fig.~\ref{fig:Fig1_0}. Ideally, its optical transfer function should follow Eq.~(\ref{eq:LOTF}). Additionally, this response should be sustained over a range of photon wavelengths, that is required to perform the imaging scanning in the $y$-direction as illustrated in Figs.~\ref{fig:Concept}c,d. This is a highly nontrivial requirement, in contrast to previous structures that were optimised for single-wavelength operation under classical continuous laser illumination~\cite{Wesemann:2021-98:LSA}.

We design the metasurface in the form of a periodic Si grating on a Si waveguide layer above a SiO$_2$ substrate. 
Part of the incident light excites waveguide modes that propagate laterally, interfere, and re-radiate, while another portion is directly transmitted. The interference between the diffracted and directly transmitted field allows one to create a tailored transmission spectrum in frequency and spatial momentum~\cite{harvey2019Opt.Eng}. 
Given the target resonant photon wavelength $\lambda$, the grating period can be approximately found as $d=\lambda/n_{e}$, where  $n_e$ is the effective mode index, and we consider first-order resonance with the normally incident plane wave.
As mentioned above, the Si metasurface must also operate 
over the photon bandwidth to cover all the scanning wavelengths (1560 to 1580~nm), while simultaneously providing a linear transfer function with sufficient slope for high sensitivity with a wide enough range $k_{t} - \Delta k_z/2 < k_z < k_{t} + \Delta k_z/2$ to maximise the spatial resolution, where $k_t$ corresponds to a fixed tilting angle of the metasurface. This requires a certain balance between the quality factor ($Q_{\rm Si}$) and the mode dispersion, which can be achieved by adjusting the
grating duty cycle $l$, since $Q_{\rm Si} \propto 1/\sin^2(l\pi)$.  We therefore performed parametric sweeps over $l$ to optimize the design. With a 100~nm thick Si thin film and 120~nm grating height, a grating period 615~nm and a duty cycle of~0.45 results in a quality factor $Q_{\rm Si}\simeq 20$. 
More details on the design procedure are provided in Sec. S1.2 of the Supplementary material.
In this design, the tilting angle of the metasurface is $\theta_t = 4^\circ$.
Most importantly, the combined effect of the grating and thin film 
allows the OTF to retain a linear form over the tunable wavelength range. Indeed, in the range $\Delta k_z =0.05$, the slope of the amplitude OTF for all three photon wavelengths shown in the shaded part of Fig.~\ref{fig:Concept}e (from 1560 to 1580~nm) is very similar, $a \simeq 11.6,~12,~12.8$ (unitless in $k_z/k_0$ plot, where $k_0=2\pi/\lambda$ at the central photon wavelength). The constant term is, 
for a tilted metasurface, $b \simeq 0.4,~0.5, ~0.7$. The wavelength dependence of $a$ and $b$ can be fully taken into account, allowing for accurate phase gradient reconstruction, as we discuss in the following. We note that the Si metasurface also introduces a phase modulation on the transmitted light, as shown with the dashed lines in Fig.~\ref{fig:Concept}e, however it does not affect the photon counts and can be therefore disregarded in the current imaging scheme.


\subsection{Experimental characterization of the metasurfaces}

\begin{figure}[h!]
    \centering
    \includegraphics[width=0.9\linewidth]{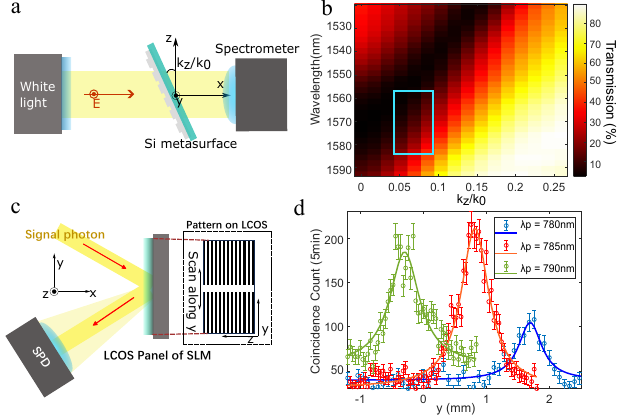}
    \caption{\textbf{a}. A transmission setup used to experimentally check the incident-angle dependent transmission of the Si metasurface by rotating it around the $y$-axis. The incident wave is $y$-polarised.  ~\textbf{b}. The experimentally measured transmission of the Si metasurface. 
    The blue box marks the applicable region for phase-gradient imaging.  ~\textbf{c}. Setup for calibrating the signal photon positions along the $y$ direction at different pump wavelengths. 
    The photon is projected on the LCOS panel of the SLM, where a pattern is formed with an unmodulated part in the centre and highly reflective sections on the sides, such that only the photons incident on the central strip are registered by a single photon detector (SPD).
    ~\textbf{d}. Experimentally measured coincidence counts between signal and idler photons over 5-minute intervals vs. the pattern position on the LCOS panel at the three pump wavelengths, revealing the distribution of signal photon positions.
    }
    \label{fig:Setup}
\end{figure}

We now perform the separate experimental characterization of linear and nonlinear metasurfaces, before their integration in a phase-gradient imaging system.


By design, a linear transmission through the Si metasurface should depend on the incident wavevector along the $z$-direction. 
%
%
This dispersion is measured through a transmission setup with $y$-polarised white light and the Si metasurface being mechanically rotated, as sketched in Fig.~\ref{fig:Setup}a. We indeed observe a linear OTF in the marked blue box of the experimentally measured transmission in Fig.~\ref{fig:Setup}b. It corresponds to the shaded region in the theoretically calculated Fig.~\ref{fig:Concept}e, which is suitable for phase-gradient imaging.


Next, we determine the spatial coordinates along the $y$-direction of the signal photons emitted from the LiNbO$_3$ metasurface, at the three selected pump wavelengths of 780, 785, and 790~nm, which were theoretically modeled in Figs.~\ref{fig:Concept}c-d.
For this purpose, we project the signal photons on the liquid crystal on silicon (LCOS) panel, and employ a hybrid test phase-modulation pattern that comprises two distinct regions, as sketched in Fig.~\ref{fig:Setup}c. One is a non-diffractive central section spanning around 0.6~mm in the $y$-direction, which preserves the incident beam wavefront. On the sides are periodic patterns that impose strong spatial phase modulation, resulting in the photon diffraction away from the single-photon detector. By shifting the pattern along the $y$-axis using the HOLOEYE Pattern Generator software, the spatial overlap between the non-diffractive central section and the transverse profile of the signal photon was systematically varied, and we recorded the coincidence count rate between the signal and idler photons. A maximum in the coincidence count occurs when the non-diffractive region aligns with the center of the signal photon beam, as this configuration minimises diffraction-induced scattering. Consequently, the $y$-coordinate of the beam center for each wavelength was identified by tracking the peak in the coincidence count rate as a function of the pattern’s positional shift. The results presented in Fig.~\ref{fig:Setup}d show that the signal photons have spatially resolved positions in the $y$-direction. The photon distributions are wider compared to the theoretical prediction in Fig.~\ref{fig:Concept}c, resulting in a small overlap, this still allows for the accurate determination of independent phase-gradients along $z$ at these locations, as we demonstrate in the following. Whereas we experimentally realise 3-pixel resolution in the $y$-direction, corresponding to the three pump wavelengths, we discuss in the following section the potential for significantly enhanced imaging resolution.

\subsection{Theoretical analysis of the phase-gradient imaging resolution} \label{sec:resolution}


\begin{figure}[!b]
    \centering
    \includegraphics[width=1\linewidth]{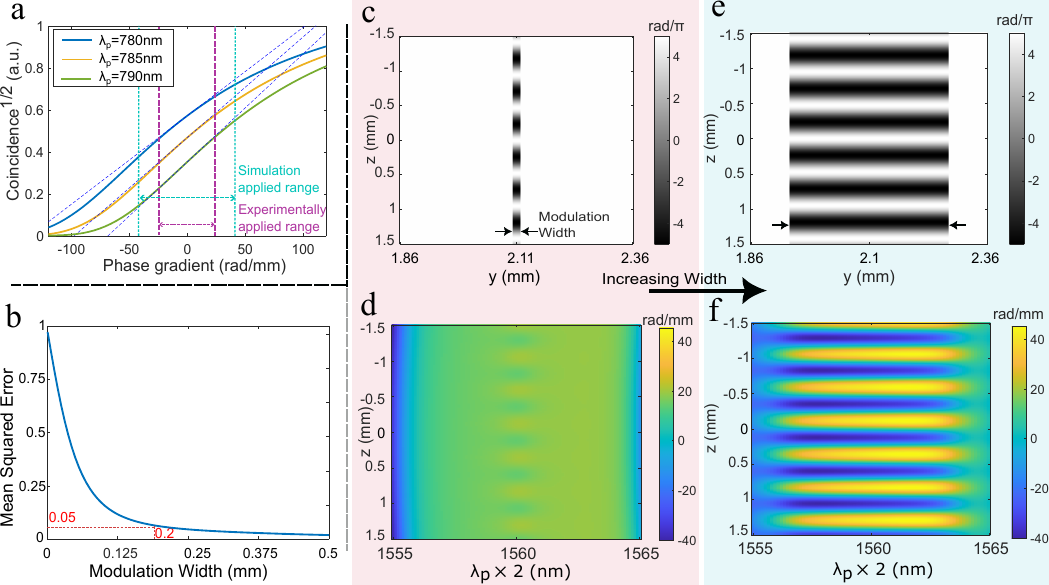}
    \caption{Resolution of quantum phase-gradient imaging. \textbf{a}.~The simulated relation between the square root of photon coincidence rate and the phase gradient $\partial\phi/\partial z$ at different pump wavelengths. The coincidences are normalised with respect to the coincidence rate without the object and the Si-metasurface. The dashed lines are the linear fitting around the zero phase gradient. 
    Labels indicate the maximum resolvable phase gradients considered in theoretical simulations and realised in experiments.
    ~\textbf{b}. The average deviation of the simulated reconstructed phase gradient from the target phase gradient at $2\lambda_p=1560$~nm vs. the phase modulation width along $y$.
    The dashed line marks the 5\% error relative to the target phase gradient, corresponding to the modulation width of 0.2~mm.  ~\textbf{c,e}. The input cosine phase patterns with the modulation width of  (c)~0.02~mm and (e)~0.35~mm. \textbf{d,f}. The phase gradient reconstruction simulated by scanning through $2\lambda_p = 1555$ to $1565$~nm for the patterns shown in (c) and (e), respectively.}
            \label{fig:yResolution}
\end{figure}

Using the designed metasurfaces and the simulated optical transfer function (OTF), we theoretically analyze the imaging resolution. 
Since the linear part of the OTF spans a limited $k_z$-space region, as shown in the shaded part of Fig.~\ref{fig:Concept}e, the metasurface can accurately resolve only a specific range of phase gradients. 
%
Under our experimental conditions,  the photon wavefunction width is $\sigma_z \simeq 87\mu m$, and we find that the term $a^2/(4 \sigma_z^2)$ can be neglected as it is much smaller than the second bracket on the right of Eq.~(\ref{eq:Demo}), see details in the Supplementary Sec.~S2.2.
Then, we perform the exact calculations based on the Fourier transforms according to Eq.~(\ref{eq:LOTF}), and analyse whether the square root of photon counts matches the theoretically expected linear dependence on $(a_{\lambda_c} \partial \phi / \partial z + b_{\lambda_c})$, where $\lambda_c$ is the central photon wavelength. The results presented in Fig.~\ref{fig:yResolution}a
indicate that the designed metasurface can resolve a maximum phase gradient of approximately $\pm$40 rad/mm across $1560-1580$~nm wavelength range when the metasurface is tilted by $4^\circ$ or $k_z/k_0 \simeq 0.07$, and $\pm$25 rad/mm with high precision of linear correspondence for any wavelength within this range. 
%
On the other hand, the maximum number of spatially resolvable object pixels along the $z$-direction is governed by the general principles of ghost imaging, being approximately proportional to the pump beam diameter and the associated nonlinear metasurface dimensions~\cite{Ma:2025-2:ELI}.

The spatial resolution of phase-gradient imaging along the $y$-direction, analogous to amplitude imaging, is directly governed by the width of spatial emission across the lithium niobate metasurface grating~\cite{Ma:2025-2:ELI}, 
which is inversely proportional to the quality factor $Q_{\rm LN}$. 
Consequently, the overall resolution of the system can be significantly enhanced by increasing the nonlinear metasurface size and the quality factor of nonlocal resonances at the photon wavelengths. 

We aim to extract the phase gradient along the $z$ direction. However, since the measurements are effectively spatially averaged over the photon wavefunction, it is important to check whether the gradient reconstruction is affected by phase variations in the orthogonal $y$-direction.
To quantify this aspect,
we perform numerical simulations for the photon emission spectrum centered at 1560~nm. We consider a phase modulation with a cosine profile along the $z$-direction with a width $w_m$ across the $y$-coordinate, while the outside area has no phase modulation, as shown in Figs.~\ref{fig:yResolution}c,e. 
We then use Eq.~(\ref{eq:Demo}), neglecting the vanishingly small term $a^2/(4 \sigma_z^2)$ as discussed above, with parameters $a$ and $b$ depending on the central photon wavelength $\lambda_c$ and the corresponding photon position $y_c$, to model the phase gradient reconstruction:
\begin{equation} \label{eq:Freconstr}
    F(z, y_c) = \left. \frac{\partial \phi}{\partial z}\right|_{y_c} = \frac{1}{a_{\lambda_c}} \left( {\sqrt{C(z, y_c)}-b_{\lambda_c} } \right) ,
\end{equation}
The calculated example phase gradient patterns by the combined ghost imaging in $z$ and scanning in $y$ direction are shown in Figs.~\ref{fig:yResolution}d,f. 
We evaluated the difference between the numerically reconstructed phase gradient $F(z,y_c)$ at the $y_c$ coordinate in the middle of the cosine pattern with the width $w_m$ and the designed phase gradient $F_t(z,y_c)$ by the mean squared error,
%
$\int|F_t(z,y_c)-F(z,y_c)|^2dz / \int|F_t(z,y_c)|^2dz$.
%
The results in Fig.~\ref{fig:yResolution}b demonstrate a decrease in the overall error as the modulation width $w_m$ increases.

Notably, the phase gradient reconstruction accuracy is above $95\%$ when the minimum feature size along the $y$-direction is greater than 0.2~mm. This width corresponds to a shift of the signal photon position when the central photon wavelength is varied by approximately 3~nm. We recall that the photon bandwidth is about 3~nm, which explains the origin of the resolution limit. Similar to amplitude imaging~\cite{Ma:2025-2:ELI}, higher resolution in $y$-direction would be reached with photon pairs having narrower spectra, which can be achieved by increasing the dimensions of the lithium niobate metasurface and its quality factor $Q_{\rm LN}$.

\subsection{Experimental quantum phase ghost imaging}

For a proof-of-principle experimental demonstration, where the resolution is primarily limited by the fabricated metasurface dimensions as discussed above, we consider objects with 6$\times$3-pixel phase-gradient patterns. The patterns were projected onto the LCOS panel, where each pixel has a size $\Delta y\times\Delta z=800\mu{\rm m} \times 350\mu$m.
Photon coincidence data for each pixel were accumulated over an 1-hour integration period, with measurements recorded at 5-minute intervals to yield 12 data points per pixel. We performed the error analysis and confirmed that the photon correlation measurements are shot-noise limited, see Supplementary Sec.~S4. We analysed T-shaped (Fig.~\ref{fig:ExpCalibration}a) and S-shaped (Fig.~\ref{fig:ExpReconst}a) patterns that are formed by sections of constant phase-gradient along $z$-direction in each pixel. The maximum phase gradient was set at 25~rad/mm, consistent with the calculated upper limit of resolvable phase gradients, as discussed above and indicated in Fig.~\ref{fig:yResolution}a. 

We find that the experimental photon collection efficiency decreases towards the pattern edges, away from the image center, see Supplementary Fig.~S3. This happens because the pattern size approaches a limit defined by the collection angle of lenses and single-photon detectors. 
%
We approximate the experimentally detected coincidence function as
%
\begin{equation}
    \widetilde{C}(z, y_c) = G(z, y_c)C(z, y_c)
\end{equation}
where $y_c$ is the coordinate corresponding to the central photon wavelength $\lambda_c = 2\lambda_p$ at a particular pump wavelength $\lambda_p$, $C(z, y_c)$ is the theoretical model of photon counts given by Eq.~(\ref{eq:Demo}), and $G(z,y_c)$ is a Gaussian function in $z$-direction with the central position $\gamma_{\lambda_c}$  and width $\beta_{\lambda_c}$:
%
\begin{equation}\label{eq:GLfitting}
    G(z, y_c) =  \exp\left(-\frac{(z-\gamma_{\lambda_c})^2}{ \beta_{\lambda_c}^2}\right).
\end{equation}
%
%
Then, we can determine the rescaled effective coincidences as
\begin{equation} \label{eq:Ceff}
    C_{\rm eff}(z, y_c ) = \sqrt{\frac{\widetilde{C}(z, y_c)}{G(z, y_c)}}
\end{equation}
and use those to reconstruct the phase gradient using Eq.~(\ref{eq:Freconstr}).
%
%


\begin{figure}[h!]
    \centering
    \includegraphics[width=0.9\linewidth]{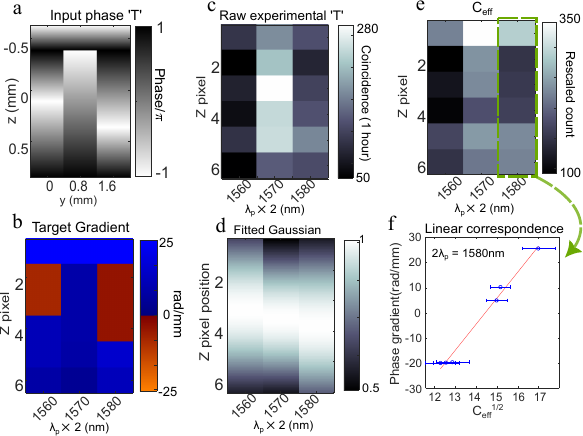}
    \caption{Calibration for phase gradient image reconstruction. \textbf{a}. The T-shaped phase object formed by multiple sections of linear phase modulation along $z$-axis. ~\textbf{b}. The corresponding calculated target phase gradient. ~\textbf{c}. The experimental shot noise limited coincidence count data of the T-shaped phase modulation pattern before processing. ~\textbf{d}. The fitted Gaussian functions $G(z)$ at the indicated central photon wavelengths. ~\textbf{e}. The rescaled coincidences $C_{\rm eff}$ according to Eq.~(\ref{eq:Ceff}).
    ~\textbf{f}. The phase gradient vs. the squared root of rescaled coincidences at $\lambda_c = 2\lambda_p=1580$~nm. Circles are experimental data with errors indicated with bars, line is a linear fitting.}
    \label{fig:ExpCalibration}
\end{figure}

We first use a single known phase-gradient pattern to determine the parameters of the Gaussian collection efficiency function $G(z, y_c)$, which can then be applied to reconstruct arbitrary objects.
Specifically, we employ a T-shaped object (Fig.~\ref{fig:ExpCalibration}a) and determine the parameters $a_{\lambda_c}$, $b_{\lambda_c}$, $\beta_{\lambda_c}$, $\gamma_{\lambda_c}$ that provide the best fitting between the theoretical phase gradient target $F_t(z)$  (Fig.~\ref{fig:ExpCalibration}b) and the reconstructed phase gradient from the experimentally measured coincidences (Fig.~\ref{fig:ExpCalibration}c) according to Eq.~(\ref{eq:Freconstr}) and  independently at each of the three wavelengths $\lambda_c$, see Supplementary Sec.~S3.
%
The fitted collection efficiency function $G(z,\lambda_c)$ is shown in Fig.~\ref{fig:ExpCalibration}d, and the rescaled coincidence counts according to Eq.~(\ref{eq:Ceff}) are presented in Fig.~\ref{fig:ExpCalibration}e. 
We then confirm that the object phase gradient has a linear dependence on the square root of $C_{\rm eff}$ in agreement with Eq.~(\ref{eq:Freconstr}), as illustrated in Fig.~\ref{fig:ExpCalibration}f for $\lambda_c = 2 \lambda_p = 1580$~nm. We also confirm the linear dependence for other wavelengths, see Supplementary Fig.~S6c.



Finally, we demonstrate a gradient reconstruction for a different object with an S-shaped phase profile shown in Fig.~\ref{fig:ExpReconst}a, where the numerically calculated phase gradient is shown in Fig.~\ref{fig:ExpReconst}b.
We experimentally measure the coincidences, see Fig.~\ref{fig:ExpReconst}c, and then perform the reconstruction with Eq.~(\ref{eq:Freconstr}), using the parameter values determined from the calibration of T-object as discussed above, and show the final result in Fig.~\ref{fig:ExpReconst}d. We see that this experimental reconstruction closely agrees with the theoretical target in Fig.~\ref{fig:ExpReconst}b, including the sign and magnitude of the phase gradient. Indeed, we calculate the overlap between Figs.~\ref{fig:ExpReconst}b,d 
and find that the reconstruction image similarity reaches 89\% (see Supplementary Table~S2). 

We further confirm the robustness of our phase-gradient reconstruction approach by swapping the phase objects. Specifically, we performed calibration using the S-shaped object, and then experimentally determined the phase gradient of the T-shaped object, reaching 82\% similarity with the theoretical reference values, see Supplementary Sec.~S3.

The experimental accuracy of the phase gradient reconstruction, which is 82\% or higher in all the cases as summarised in Supplementary Table~S2, was primarily limited by the shot noise. It can therefore be improved by using more efficient single-photon detectors and optimising the nonlinear metasurface to increase the photon-pair generation rate.


\begin{figure}[h!]
    \centering
    \includegraphics[width=1\linewidth]{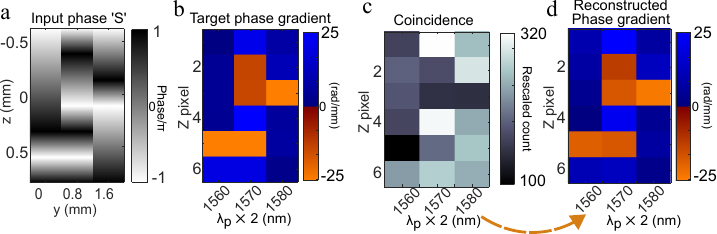}
    \caption{Proof-of-principle demonstration of quantum phase-gradient imaging. \textbf{a}. The input S-shaped phase modulation. ~\textbf{b}. The theoretical phase gradient of the S pattern ~\textbf{c}. The rescaled coincidence data $C_{\rm eff}$ calculated from raw experimental coincidences using Eq.~(\ref{eq:Ceff}) with the $G$ function from Fig.~\ref{fig:ExpCalibration}d. ~\textbf{d}. The reconstructed phase gradient of S-shaped phase pattern using Eq.~(\ref{eq:Freconstr}) with $C_{\rm eff}$ from plot (c) and $a_{\lambda_c}$, $b_{\lambda_c}$ determined from the T-object calibration in Fig.~\ref{fig:ExpCalibration}f and Supplementary Fig.~S6c.}
    \label{fig:ExpReconst}
\end{figure}

\section{Discussion}

By realising the two key functions of photon generation and phase gradient extraction with metasurfaces,
the system achieves a level of compactness and integration that was out of reach in previous quantum phase imaging setups. This dual-role metasurface method not only shrinks the physical size to the millimetre scale but also alleviates the challenges of alignment and associated stability in multi-element bulk-optical systems.

A key strength of our configuration is its large field of view combined with all-optical tunability in quantum light generation. This tunability, along with the spatial mode control provided by the nonlocal resonances in LiNbO$_3$ metasurface, can allow the system to adapt to various imaging targets and operational conditions, ranging from biological specimens to larger structured objects. 

There is a large scope for increasing the imaging performance beyond the current proof-of-principle experiment towards future applications.
First, enlarging the dimensions of the LiNbO$_3$ metasurface would allow it to tailor higher spatial frequency components of the two-photon field. Second, shaping the pump beam so that it is flatter and more uniform across the metasurface aperture would further enhance performance. Both adjustments broaden the correlation bandwidth in momentum space, which ultimately sharpens the precision of axial and transverse phase gradient measurements.

A particularly notable aspect of the developed protocol is that it achieves phase gradient extraction without the need for an interferometer or any direct interference measurements. Instead, the phase information is embedded in engineered two-photon correlations, shaped entirely by the metasurface. This design eliminates the stringent stability requirements and vulnerability to environmental noise that often constrain interferometer-based quantum phase imaging, resulting in a simpler experimental setup and greater robustness for real-world applications.

Furthermore, we demonstrate that positioning the metasurface for phase gradient extraction directly after the phase object still delivers phase retrieval precision well within acceptable limits for most practical uses. This means that relay optics or telescopic beam expansion are not strictly required, further enhancing the system’s compactness. For general-purpose imaging, the capability to perform quantum phase measurements without long free-space propagation stages marks an important step toward truly portable and field-deployable quantum sensors.

Moreover, there is an interesting prospect of performing selective extraction of gradients along different directions in the object plane by generating photon pairs with hyperentanglement in spatial and polarization degrees of freedom, and designing the Si metasurface for polarization-dependent direction of phase differentiation.

Taken together, the system shows that integrating metasurfaces can fundamentally expand the design possibilities for quantum imaging systems. By uniting photon-pair generation, state manipulation, and phase retrieval within a single class of nanostructured components, this approach paves the way for scalable, multi-functional quantum optical devices - free from the size and complexity constraints of traditional bulk optics.

\section{Conclusion}

In this study, we have developed and realised experimentally a compact and versatile quantum phase-gradient imaging system that combines a
LiNbO$_3$ metasurface for generating spatially entangled photon pairs
and a silicon (Si) metasurface for phase gradient extraction, marking the first demonstration of metasurfaces being used for both generation and detection in a quantum imaging system.

The compact design of the system is a key innovation, enabled by the use of nonlocal resonances in both metasurfaces.
The Si metasurface, in particular, plays a critical role in extracting phase gradients through its linear optical transfer function (OTF), allowing for the reconstruction of phase information with spatial quantum correlation. This approach not only simplifies the optical setup but also enhances the system's portability, making it suitable for applications where space and efficiency are paramount. Another notable feature of the system is its switchable operation between quantum phase-gradient imaging with the Si metasurface and amplitude imaging without the Si metasurface.

Our experimental results validate the theoretical framework, showing that the system can resolve phase gradients up to 25 rad/mm with high fidelity. The proof-of-principle experiment, using a calibration-based reconstruction method, achieves a similarity of 89$\%$ with the target phase gradient. This lays the groundwork for further optimisation, including improvements in resolution and contrast through large-area LiNbO$_3$ metasurface fabrication and enhancement of the resonance quality factor.

In conclusion, this work establishes a new paradigm for compact and switchable quantum phase-gradient imaging, leveraging the dual functionality of metasurfaces. The integration of generation and detection within a single platform opens up exciting possibilities for applications in quantum imaging, sensing, biomedical metrology, and beyond. In particular, all-optical scanning of the photon emission direction can underpin LiDAR-like fast tracking and imaging of transparent phase-only objects in complex and dynamic environments.

\section{Methods}

The experimental setup is detailed as follows. A tunable pump laser for spontaneous parametric down-conversion (FPL785P, Thorlabs) operates within a wavelength range of 779 to 791~nm. The laser beam is passed through a 200~mm focal length lens and an 850~nm short-pass filter before being projected onto the LiNbO$_3$ metasurface. The beam is focused to a waist of 200~$\mu$m on the metasurface, which is mounted on a 3D translation stage for precise positioning. A 50~mm lens is then used to collimate the emitted photons.

The photons generated through SPDC are initially $z$-polarized. Since the z-oriented optical transfer function supported by the Si metasurface is functional for only the $y$-polarized light, a half-wave plate oriented at 45 degrees is introduced to rotate the polarization of the photon pairs generated from the LiNbO$_3$ metasurface. 
Both photons are passed through a long pass filter at 1100 nm and a band-pass filter centered at 1570~nm with a bandwidth of $\pm$50~nm to suppress the pump, and then a D-shaped mirror is used to split the signal and idler photons.

One optical path includes a scanning slit moving along $z$, which mimics a 1D detector array, while the other path incorporates an SLM (PLUTO 2.1, HOLOEYE) to produce the phase pattern, followed by the Si metasurface. The orientation of the SLM panel and the projected phase pattern is aligned with the signal photon polarization. The Si metasurface and the scanning slit are positioned at equal distances from the D-shaped mirror to ensure high-precision correlation measurements. 
The photons are directed to single-photon detectors based on InGaAs/InP avalanche photodiodes (ID230, ID Quantique) via multimode fibers. 

Coincidence events are characterized using a time-to-digital converter (ID801, ID Quantique), with a coincidence window set to 0.42~ns for precise timing correlation analysis.

\section{Fabrication}

The 400~$\mu$m $\times$ 400~$\mu m$ LiNbO$_3$ metasurface was fabricated on a 303.7~nm thin film on a quartz substrate (NANOLN). The fabrication process began with depositing a 180 nm SiO$_2$ layer via plasma-enhanced chemical vapour deposition. A nickel hard mask was then patterned using electron beam lithography with a PMMA resist and a lift-off process. This mask was used to etch the SiO$_2$ layer with inductively coupled plasma, after which the nickel was chemically removed~\cite{Ma:2025-2:ELI}.

The silicon metasurface was fabricated using electron beam lithography (EBL) combined with reactive ion etching (RIE). First, commercially available amorphous silicon thin films on glass substrates (Tafelmaier Dünnschicht Technik GmbH) were cleaned in a specialised machine (OPTIwet SB~30). The silicon thickness was adjusted to the target value by argon-ion beam etching, followed by the deposition of a conductive chromium layer and a 100~nm negative electron-beam resist (EN038, Tokyo Ohka Kogyo Co., Ltd.). The resist was then patterned using a variable-shaped beam electron-beam lithography system (Vistec SB 350).

After exposure, the resist was developed in a developer (OPD 4262), and the pattern was transferred to the chromium layer by ion beam etching (Oxford Ionfab 300). The underlying silicon layer was then etched to the desired trench depth using reactive ion etching (RIE-ICP, Sentech SI-500 C). Finally, the residual resist and chromium layer were removed in acetone and a ceric ammonium nitrate-based solution, respectively.

\section{Acknowledgment}
This work was supported by the Australian Research Council (\url{https://doi.org/10.13039/501100000923}) Centre of Excellence for Transformative Meta-Optical Systems - TMOS (CE200100010) and the Deutsche Forschungsgemeinschaft (DFG, German Research Foundation) project Meta Active IRTG 2675 (437527638).

\section{Disclosures}
The authors declare no conflicts of interest.

\printbibliography

@STRING{ACS = "Acoust. Phys."}

@STRING{APR = "Appl. Phys. Rev."}

@STRING{ARXIV = "arXiv"}

@STRING{CPH = "Commun. Phys."}

@STRING{ELL = "Electron. Lett."}

@STRING{LSA = "Light Sci. Appl."}

@STRING{NANL = "Nano Lett."}

@STRING{NANP = "Nanophotonics"}

@STRING{NAT = "Nature"}

@STRING{NPHOT = "Nat. Photon."}

@STRING{PRA = "Phys. Rev. A"}

@STRING{PT = "Phys. Today"}

@STRING{PTRSA = "Philos. Trans. R. Soc. A"}

@STRING{QST = "Quantum Sci. Technol."}

@STRING{SCA = "Sci. Adv."}

@STRING{SCI = "Science"}

@Article{Zhang:2022-eabq4240:SCA,
	author	    = {Zhang, J. H. and Ma, J. Y. and Parry, M. and
	  Cai, M. and Camacho-Morales, R. and Xu, L. and Neshev, D. N.
	  and Sukhorukov, A. A.},
	title	    = {Spatially entangled photon pairs from lithium
	  niobate nonlocal metasurfaces},
	journal     = SCA,
	year	    = {2022},
	month	     = {Jul 29},
	volume	    = {8},
	number	    = {30},
	pages	    = {eabq4240},
	database      = {WOS:000836554300040},
	citetimes      = {0},
	language      = {English},
	doi	 = {10.1126/sciadv.abq4240},
        url      = {https://doi.org/10.1126/sciadv.abq4240}
}

@Article{Padgett:2017-20160233:PTRSA,
	author	    = {Padgett, M. J. and Boyd, R. W.},
	title	    = {An introduction to ghost imaging: quantum and
	  classical},
	journal     = PTRSA,
	year	    = {2017},
	month	     = {Aug 6},
	volume	    = {375},
	number	    = {2099},
	pages	    = {20160233},
	database      = {WOS:000412172300001},
	citetimes      = {77},
	language      = {English},
	doi	 = {10.1098/rsta.2016.0233},
        url      = {https://doi.org/10.1098/rsta.2016.0233}
}

@Article{Santiago-Cruz:2022-991:SCI,
	author	    = {Santiago-Cruz, T. and Gennaro, S. D. and
	  Mitrofanov, O. and Addamane, S. and Reno, J. and Brener, I.
	  and Chekhova, M. V.},
	title	    = {Resonant metasurfaces for generating complex
	  quantum states},
	journal     = SCI,
	year	    = {2022},
	month	     = {Aug 26},
	volume	    = {377},
	number	    = {6609},
	pages	    = {991-995},
	database      = {WOS:000849788200037},
	citetimes      = {0},
	language      = {English},
	doi	 = {10.1126/science.abq8684},
        url      = {https://doi.org/10.1126/science.abq8684}
}

@Article{Santiago-Cruz:2021-4423:NANL,
	author	    = {Santiago-Cruz, T. and Fedotova, A. and Sultanov,
	  V. and Weissflog, M. A. and Arslan, D. and Younesi, M. and
	  Pertsch, T. and Staude, I. and Setzpfandt, F. and Chekhova,
	  M.},
	title	    = {Photon Pairs from Resonant Metasurfaces},
	journal     = NANL,
	year	    = {2021},
	month	     = {May 26},
	volume	    = {21},
	number	    = {10},
	pages	    = {4423-4429},
	database      = {WOS:000657242300039},
	citetimes      = {15},
	language      = {English},
	doi	 = {10.1021/acs.nanolett.1c01125},
        url      = {https://doi.org/10.1021/acs.nanolett.1c01125}
}

@STRING{ELI = "eLight"}

@STRING{JPPH = "J. Phys. Phot."}

@article{Zhu2017,
   author = {Tengfeng Zhu and Yihan Zhou and Yijie Lou and Hui Ye and Min Qiu and Zhichao Ruan and Shanhui Fan},
   doi = {10.1038/ncomms15391},
   issn = {2041-1723},
   issue = {1},
   journal = {Nat Commun},
   month = {5},
   pages = {15391},
   title = {Plasmonic computing of spatial differentiation},
   volume = {8},
   url = {https://www.nature.com/articles/ncomms15391},
   year = {2017},
}

@article{Wesemann:APR:2021,
   author = {Lukas Wesemann and Timothy J Davis and Ann Roberts},
   doi = {10.1063/5.0048758},
   issn = {1931-9401},
   issue = {3},
   journal = {Applied Physics Reviews},
   month = {9},
   pages = {31309},
   title = {Meta-optical and thin film devices for all-optical information processing},
   volume = {8},
   url = {https://pubs.aip.org/aip/apr/article/124804},
   year = {2021},
}

@article{Nguyen2022,
   author = {Thang L Nguyen and Soorya Pradeep and Robert L Judson-Torres and Jason Reed and Michael A Teitell and Thomas A Zangle},
   doi = {10.1021/acsnano.1c11507},
   issn = {1936-0851, 1936-086X},
   issue = {8},
   journal = {ACS Nano},
   month = {8},
   pages = {11516-11544},
   title = {Quantitative Phase Imaging: Recent Advances and Expanding Potential in Biomedicine},
   volume = {16},
   url = {https://pubs.acs.org/doi/10.1021/acsnano.1c11507},
   year = {2022},
}

@article{Fienup2013,
   author = {James R Fienup},
   doi = {10.1364/AO.52.000045},
   issn = {1559-128X, 2155-3165},
   issue = {1},
   journal = {Appl. Opt.},
   month = {1},
   pages = {45},
   title = {Phase retrieval algorithms: a personal tour [Invited]},
   volume = {52},
   url = {https://opg.optica.org/abstract.cfm?URI=ao-52-1-45},
   year = {2013},
}

@article{Ji2022,
   author = {Anqi Ji and Jung-Hwan Song and Qitong Li and Fenghao Xu and Ching-Ting Tsai and Richard C Tiberio and Bianxiao Cui and Philippe Lalanne and Pieter G Kik and David A B Miller and Mark L Brongersma},
   doi = {10.1038/s41467-022-34197-6},
   issn = {2041-1723},
   issue = {1},
   journal = {Nat Commun},
   month = {12},
   pages = {7848},
   title = {Quantitative phase contrast imaging with a nonlocal angle-selective metasurface},
   volume = {13},
   url = {https://www.nature.com/articles/s41467-022-34197-6},
   year = {2022},
}

@Article{Wesemann:2021-98:LSA,
	author	    = {Wesemann, L. and Rickett, J. and Song, J. C. and
	  Lou, J. Q. and Hinde, E. and Davis, T. J. and Roberts, A.},
	title	    = {Nanophotonics enhanced coverslip for phase
	  imaging in biology},
	journal     = LSA,
	year	    = {2021},
	month	     = {May 8},
	volume	    = {10},
	number	    = {1},
	pages	    = {98},
	database      = {WOS:000656445800001},
	citetimes      = {3},
	language      = {English},
	doi	 = {10.1038/s41377-021-00540-7},
        url      = {https://doi.org/10.1038/s41377-021-00540-7}
}

@article{Silva:Science:2014,
author = {Alexandre Silva  and Francesco Monticone  and Giuseppe Castaldi  and Vincenzo Galdi  and Andrea Alù  and Nader Engheta },
title = {Performing Mathematical Operations with Metamaterials },
journal = {Science},
volume = {343},
number = {6167},
pages = {160-163},
year = {2014},
doi = {10.1126/science.1242818},
URL = {https://www.science.org/doi/abs/10.1126/science.1242818},
eprint = {https://www.science.org/doi/pdf/10.1126/science.1242818}}

@article{PhysRevD.23.1693,
  title = {Quantum-mechanical noise in an interferometer},
  author = {Caves, Carlton M.},
  journal = {Phys. Rev. D},
  volume = {23},
  issue = {8},
  pages = {1693--1708},
  numpages = {0},
  year = {1981},
  month = {Apr},
  publisher = {American Physical Society},
  doi = {10.1103/PhysRevD.23.1693},
  url = {https://link.aps.org/doi/10.1103/PhysRevD.23.1693}
}

@Article{Park2018,
author={Park, YongKeun
and Depeursinge, Christian
and Popescu, Gabriel},
title={Quantitative phase imaging in biomedicine},
journal={Nature Photonics},
year={2018},
month={Oct},
day={01},
volume={12},
number={10},
pages={578-589},
issn={1749-4893},
doi={10.1038/s41566-018-0253-x},
url={https://doi.org/10.1038/s41566-018-0253-x}
}

@Article{Ortolano2023,
author={Ortolano, Giuseppe
and Paniate, Alberto
and Boucher, Pauline
and Napoli, Carmine
and Soman, Sarika
and Pereira, Silvania F.
and Ruo-Berchera, Ivano
and Genovese, Marco},
title={Quantum enhanced non-interferometric quantitative phase imaging},
journal={Light: Science {\&} Applications},
year={2023},
month={Jul},
day={11},
volume={12},
number={1},
pages={171},
issn={2047-7538},
doi={10.1038/s41377-023-01215-1},
url={https://doi.org/10.1038/s41377-023-01215-1}
}

@article{Giovannetti_2011,
   title={Advances in quantum metrology},
   volume={5},
   ISSN={1749-4893},
   url={http://dx.doi.org/10.1038/nphoton.2011.35},
   DOI={10.1038/nphoton.2011.35},
   number={4},
   journal={Nature Photonics},
   publisher={Springer Science and Business Media LLC},
   author={Giovannetti, Vittorio and Lloyd, Seth and Maccone, Lorenzo},
   year={2011},
   month=mar, pages={222–229} }

@article{ZERNIKE1942974,
title = {Phase contrast, a new method for the microscopic observation of transparent objects part II},
journal = {Physica},
volume = {9},
number = {10},
pages = {974-986},
year = {1942},
issn = {0031-8914},
doi = {https://doi.org/10.1016/S0031-8914(42)80079-8},
url = {https://www.sciencedirect.com/science/article/pii/S0031891442800798},
author = {F. Zernike}
}

@article{Bykov:2018：optica,
author = {Dmitry A. Bykov and Leonid L. Doskolovich and Andrey A. Morozov and Vladimir V. Podlipnov and Evgeni A. Bezus and Payal Verma and Victor A. Soifer},
journal = {Opt. Express},
number = {8},
pages = {10997--11006},
publisher = {Optica Publishing Group},
title = {First-order optical spatial differentiator based on a guided-mode resonant grating},
volume = {26},
month = {Apr},
year = {2018},
url = {https://opg.optica.org/oe/abstract.cfm?URI=oe-26-8-10997},
doi = {10.1364/OE.26.010997},
}

@article{Wang:2020:ACSPh,
author = {Wang, Haiwen and Guo, Cheng and Zhao, Zhexin and Fan, Shanhui},
title = {Compact Incoherent Image Differentiation with Nanophotonic Structures},
journal = {ACS Photonics},
volume = {7},
number = {2},
pages = {338-343},
year = {2020},
doi = {10.1021/acsphotonics.9b01465}
}

@misc{rusca2024QKDarxiv,
      title={Quantum Cryptography: an overview of Quantum Key Distribution}, 
      author={Davide Rusca and Nicolas Gisin},
      year={2024},
      eprint={2411.04044},
      archivePrefix={arXiv},
      primaryClass={quant-ph},
      url={https://arxiv.org/abs/2411.04044}, 
}

@Article{Kwon:NaturePhotonics:2020,
author={Kwon, Hyounghan
and Arbabi, Ehsan
and Kamali, Seyedeh Mahsa
and Faraji-Dana, MohammadSadegh
and Faraon, Andrei},
title={Single-shot quantitative phase gradient microscopy using a system of multifunctional metasurfaces},
journal={Nature Photonics},
year={2020},
month={Feb},
day={01},
volume={14},
number={2},
pages={109-114},
issn={1749-4893},
doi={10.1038/s41566-019-0536-x},
url={https://doi.org/10.1038/s41566-019-0536-x}
}

@article{harvey2019Opt.Eng,
  author    = {Harvey, James E. and Pfisterer, Richard N.},
  title     = {{Understanding diffraction grating behavior: including conical diffraction and Rayleigh anomalies from transmission gratings}},
  journal   = {Optical Engineering},
  volume    = {58},
  number    = {8},
  pages     = {087105},
  year      = {2019},
  doi       = {10.1117/1.OE.58.8.087105},
}

@Article{Wang:2022-38:PT,
	author	    = {Wang, K. and Chekhova, M. and Kivshar, Y.},
	title	    = {METASURFACES for quantum technologies},
	journal     = PT,
	year	    = {2022},
	month	     = {Aug 1},
	volume	    = {75},
	number	    = {8},
	pages	    = {38-44},
	database      = {WOS:000847298300015},
	citetimes      = {3},
	language      = {English},
    doi = {10.1063/PT.3.5062}
}

@Article{Jia:2025-eads3576:SCA,
	author	    = {Jia, W. H. and Saerens, G. and Talts, UL. and
	  Weigand, H. and Chapman, R. J. and Li, L. and Grange, R. and
	  Yang, Y. M.},
	title	    = {Polarization-entangled {B}ell state generation
	  from an epsilon-near-zero metasurface},
	journal     = SCA,
	year	    = {2025},
	month	     = {Feb 21},
	volume	    = {11},
	number	    = {8},
	pages	    = {eads3576},
	database      = {WOS:001428018300019},
	citetimes      = {0},
	language      = {English},
	doi	 = {10.1126/sciadv.ads3576},
        url      = {https://doi.org/10.1126/sciadv.ads3576}
}

@article{Noh:2024-15356:NANL,
  title = {Quantum {{Pair Generation}} in {{Nonlinear Metasurfaces}} with {{Mixed}} and {{Pure Photon Polarizations}}},
  author = {Noh, Jiho and {Santiago-Cruz}, Tom{\'a}s and Sultanov, Vitaliy and Doiron, Chloe F. and Gennaro, Sylvain D. and Chekhova, Maria V. and Brener, Igal},
  year = {2024},
  month = dec,
  journal = NANL,
  volume = {24},
  number = {48},
  pages = {15356--15362},
  issn = {1530-6984, 1530-6992},
  doi = {10.1021/acs.nanolett.4c04398},
  urldate = {2025-03-14},
  copyright = {https://doi.org/10.15223/policy-029},
  langid = {english}
}

@Article{Ma:2023-8091:NANL,
	author	    = {Ma, J. Y. and Zhang, J. H. and Jiang, Y. X. and
	  Fan, T. M. and Parry, M. and Neshev, D. N. and Sukhorukov, A.
	  A.},
	title	    = {Polarization Engineering of Entangled Photons
	  from a Lithium Niobate Nonlinear Metasurface},
	journal     = NANL,
	year	    = {2023},
	month	     = {Aug 23},
	volume	    = {23},
	number	    = {17},
	pages	    = {8091-8098},
	database      = {WOS:001063458100001},
	citetimes      = {14},
	language      = {English},
	doi	 = {10.1021/acs.nanolett.3c02055},
        url      = {https://doi.org/10.1021/acs.nanolett.3c02055}
}

@Article{Ma:2025-2:ELI,
	author	    = {Ma, J. Y. and Ren, J. L. and Zhang, J. H. and
	  Meng, J. J. and McManus-Barrett, C. and Crozier, K. B. and
	  Sukhorukov, A. A.},
	title	    = {Quantum imaging using spatially entangled photon
	  pairs from a nonlinear metasurface},
	journal     = ELI,
	year	    = {2025},
	month	     = {Dec},
	volume	    = {5},
	number	    = {1},
	pages	    = {2},
	database      = {WOS:001416168200001},
	citetimes      = {0},
	language      = {English},
	doi	 = {10.1186/s43593-024-00080-8},
        url      = {https://doi.org/10.1186/s43593-024-00080-8}
}

@Article{Ma:2025-eadu4133:SCA,
	author	    = {Ma, J. Y. and Fan, T. M. and Haggren, T. and
	  Molina, L. V. and Parry, M. and Shinde, S. and
	  McManus-Barrett, C. and Zhang, J. H. and Morales, R. C. and
	  Setzpfandt, F. and Tan, H. H. and Jagadish, C. and Neshev, D.
	  N. and Sukhorukov, A. A.},
	title	    = {Nonlinearity symmetry breaking for generating
	  tunable quantum entanglement in semiconductor metasurfaces},
	journal     = SCA,
	year	    = {2025},
	month	     = {Jul 9},
	volume	    = {11},
	number	    = {28},
	pages	    = {eadu4133},
	database      = {WOS:001525278400016},
	citetimes      = {1},
	language      = {English},
	doi	 = {10.1126/sciadv.adu4133},
        url      = {https://doi.org/10.1126/sciadv.adu4133}
}

@Article{Defienne:2024-1024:NPHOT,
	author	    = {Defienne, H. and Bowen, W. P. and Chekhova, M.
	  and Lemos, G. B. and Oron, D. and Ramelow, S. and Treps, N.
	  and Faccio, D.},
	title	    = {Advances in quantum imaging},
	journal     = NPHOT,
	year	    = {2024},
	month	     = {Oct},
	volume	    = {18},
	number	    = {10},
	pages	    = {1024-1036},
	database      = {WOS:001322450800001},
	citetimes      = {53},
	language      = {English},
	doi	 = {10.1038/s41566-024-01516-w},
        url      = {https://doi.org/10.1038/s41566-024-01516-w}
}

@article{Noh:2025-371:LSA,
  title = {Fano interference of photon pairs from a metasurface},
  volume = {14},
  pages={371},
  url = {https://doi.org/10.1038/s41377-025-01998-5},
  DOI = {10.1038/s41377-025-01998-5},
  journal = LSA,
  author = {Noh,  Jiho and Santiago-Cruz,  Tomás and Doiron,  Chloe F. and Jung,  Hyunseung and Yu,  Jaeyeon and Addamane,  Sadhvikas J. and Chekhova,  Maria V. and Brener,  Igal},
  year = {2025}
}

@Article{Lemos:2014-409:NAT,
	author	    = {Lemos, G. B. and Borish, V. and Cole, G. D. and
	  Ramelow, S. and Lapkiewicz, R. and Zeilinger, A.},
	title	    = {Quantum imaging with undetected photons},
	journal     = NAT,
	year	    = {2014},
	month	     = {Aug 28},
	volume	    = {512},
	number	    = {7515},
	pages	    = {409-412},
	database      = {WOS:000340840600028},
	citetimes      = {307},
	language      = {English},
	doi	 = {10.1038/nature13586},
        url      = {https://doi.org/10.1038/nature13586}
}

@Article{Paterova:2018-25008:QST,
	author	    = {Paterova, A. V. and Yang, H. Z. and An, C. W.
	  and Kalashnikov, D. A. and Krivitsky, L. A.},
	title	    = {Tunable optical coherence tomography in the
	  infrared range using visible photons},
	journal     = QST,
	year	    = {2018},
	month	     = {Apr},
	volume	    = {3},
	number	    = {2},
	pages	    = {025008},
	database      = {WOS:000429341500001},
	citetimes      = {27},
	language      = {English},
	doi	 = {10.1088/2058-9565/aab567},
        url      = {https://doi.org/10.1088/2058-9565/aab567}
}

@Article{Forbes:2025-174:CPH,
	author	    = {Forbes, A. and Nothlawala, F.},
	title	    = {How a thirty-year-old quantum tale of two
	  photons became ghost imaging},
	journal     = CPH,
	year	    = {2025},
	month	     = {Apr 19},
	volume	    = {8},
	number	    = {1},
	pages	    = {174},
	database      = {WOS:001471182200002},
	citetimes      = {0},
	language      = {English},
	doi	 = {10.1038/s42005-025-02099-w},
        url      = {https://doi.org/10.1038/s42005-025-02099-w}
}

@Article{Pittman:1995-3429:PRA,
	author	    = {Pittman, T. B. and Shih, Y. H. and Strekalov, D.
	  V. and Sergienko, A. V.},
	title	    = {Optical imaging by means of 2-photon quantum
	  entanglement},
	journal     = PRA,
	year	    = {1995},
	month	     = {Nov},
	volume	    = {52},
	number	    = {5},
	pages	    = {R3429-R3432},
	database      = {WOS:A1995TE17300011},
	citetimes      = {1153},
	language      = {English},
	doi	 = {10.1103/PhysRevA.52.R3429},
        url      = {https://doi.org/10.1103/PhysRevA.52.R3429}
}

@Article{Cameron:2024-33001:JPPH,
	author	    = {Cameron, P. and Courme, B. and Faccio, D. and
	  Defienne, H.},
	title	    = {Shaping the spatial correlations of entangled
	  photon pairs},
	journal     = JPPH,
	year	    = {2024},
	month	     = {Jul 1},
	volume	    = {6},
	number	    = {3},
	pages	    = {033001},
	database      = {WOS:001248910500001},
	citetimes      = {0},
	language      = {English},
	doi	 = {10.1088/2515-7647/ad50b1},
        url      = {https://doi.org/10.1088/2515-7647/ad50b1}
}

@article{kanAdvances2023,
  title = {Advances in {{Metaphotonics Empowered Single Photon Emission}}},
  author = {Kan, Yinhui and Bozhevolnyi, Sergey I.},
  year = {2023},
  month = may,
  journal = {Advanced Optical Materials},
  volume = {11},
  number = {10},
  pages = {2202759},
  issn = {2195-1071, 2195-1071},
  doi = {10.1002/adom.202202759},
  NOurldate = {2023-11-28},
  langid = {english}
}

@article{maEngineering2024,
  title = {Engineering {{Quantum Light Sources}} with {{Flat Optics}}},
  author = {Ma, Jinyong and Zhang, Jihua and Horder, Jake and Sukhorukov, Andrey A. and Toth, Milos and Neshev, Dragomir N. and Aharonovich, Igor},
  year = {2024},
  month = apr,
  journal = {Advanced Materials},
  pages = {2313589},
  issn = {0935-9648, 1521-4095},
  doi = {10.1002/adma.202313589},
  NOurldate = {2024-05-15},
  langid = {english}
}

@article{gilabertebassetPerspectives2019,
	title = {Perspectives for {Applications} of {Quantum} {Imaging}},
	volume = {13},
	issn = {1863-8880, 1863-8899},
	url = {https://onlinelibrary.wiley.com/doi/10.1002/lpor.201900097},
	doi = {10.1002/lpor.201900097},
	language = {en},
	number = {10},
	NOurldate = {2022-09-27},
	journal = {Laser \& Photonics Reviews},
	author = {Gilaberte Basset, Marta and Setzpfandt, Frank and Steinlechner, Fabian and Beckert, Erik and Pertsch, Thomas and Gräfe, Markus},
	month = oct,
	year = {2019},
	pages = {1900097},
}

@article{shihQuantum2007c,
	title = {Quantum {Imaging}},
	volume = {13},
	issn = {1077-260X},
	url = {http://ieeexplore.ieee.org/document/4303053/},
	doi = {10.1109/JSTQE.2007.902724},
	language = {en},
	number = {4},
	NOurldate = {2023-10-03},
	journal = {IEEE Journal of Selected Topics in Quantum Electronics},
	author = {Shih, Yanhua},
	year = {2007},
	pages = {1016--1030},
}

@article{bridaExperimental2010a,
	title = {Experimental realization of sub-shot-noise quantum imaging},
	volume = {4},
	issn = {1749-4885, 1749-4893},
	url = {https://www.nature.com/articles/nphoton.2010.29},
	doi = {10.1038/nphoton.2010.29},
	language = {en},
	number = {4},
	NOurldate = {2023-10-03},
	journal = {Nature Photonics},
	author = {Brida, G. and Genovese, M. and Ruo Berchera, I.},
	month = apr,
	year = {2010},
	pages = {227--230},
}

@article{samantarayRealization2017,
	title = {Realization of the first sub-shot-noise wide field microscope},
	volume = {6},
	issn = {2047-7538},
	url = {https://www.nature.com/articles/lsa20175},
	doi = {10.1038/lsa.2017.5},
	language = {en},
	number = {7},
	NOurldate = {2023-10-03},
	journal = {Light: Science \& Applications},
	author = {Samantaray, Nigam and Ruo-Berchera, Ivano and Meda, Alice and Genovese, Marco},
	month = jan,
	year = {2017},
	pages = {e17005--e17005},
}

@article{moreauImaging2019,
	title = {Imaging with quantum states of light},
	volume = {1},
	issn = {2522-5820},
	url = {https://www.nature.com/articles/s42254-019-0056-0},
	doi = {10.1038/s42254-019-0056-0},
	language = {en},
	number = {6},
	NOurldate = {2024-05-17},
	journal = {Nature Reviews Physics},
	author = {Moreau, Paul-Antoine and Toninelli, Ermes and Gregory, Thomas and Padgett, Miles J.},
	month = may,
	year = {2019},
	pages = {367--380},
}

@Article{Weissflog:2024-3563:NANP,
	author	    = {Weissflog, M. A. and Ma, J. Y. and Zhang, J. H.
	  and Fan, T. M. and Lung, S. and Pertsch, T. and Neshev, D. N.
	  and Saravi, S. and Setzpfandt, F. and Sukhorukov, A. A.},
	title	    = {Directionally tunable co- and counterpropagating
	  photon pairs from a nonlinear metasurface},
	journal     = NANP,
	year	    = {2024},
	month	     = {Aug 20},
	volume	    = {13},
	number	    = {18},
	pages	    = {3563-3573},
	database      = {WOS:001252861400001},
	citetimes      = {1},
	language      = {English},
	doi	 = {10.1515/nanoph-2024-0122},
        url      = {https://doi.org/10.1515/nanoph-2024-0122}
}

\end{document}